
\input harvmac
\Title{FERMI-PUB 93/074-T}{Tachyon Condensates and String Theoretic Inflation}
\centerline{{\bf E. Raiten}\footnote{$^\dagger$}
{e-mail:\ raiten@fnth07.fnal.gov}}
\bigskip\centerline{Theory Group, MS106}
\centerline{Fermi National Accelerator Laboratory}
\centerline{P.O. Box 500, Batavia, IL 60510}

\vskip .3in
Cosmological solutions of the beta function equations for the
background fields of the closed bosonic string are
investigated at the one-loop level.  Following recent workthe of Kostelecky and
Perry, it is assumed that the spatial sections of the space-time are
conformally flat.  Working in the sigma-model frame, the non-trivial
tachyon potential is utilized to determine solutions with
sufficient inflation to solve the smoothness and flatness problems.
The graceful exit and density perturbation constraints can also
be successfully implemented.
\Date{4/93}
%
\def\pl#1{{\it Phys. Lett.}{\bf #1B}}
\def\np#1{{\it Nucl. Phys.}{\bf B#1}}

\def\prl#1{{\it Phys. Rev. Lett.}{\bf#1}}
\def\prd#1{{\it Phys. Rev.}{\bf D#1}}
\def\ijmpd#1{{\it Int. J. Mod Phys.}{\bf D#1}}
\def\p{\phi}

\def\ap{\alpha '}

\def\l{\lambda}
\lref\cfmp{C.G. Callan, D. Friedan, E.J. Martinec, M.J. Perry, \np{262}
(1985) 593.}
\lref\kt{E.W. Kolb, M.S. Turner, {\it The Early Universe}, Addison-Wesley
(Redwood City, CA), 1990.}
\lref\tsea{A.A. Tseytlin, \ijmpd{1} (1992) 223.}
\lref\tsevaf{A.A. Tseytlin, C. Vafa, \np{372} (1992) 443.}
\lref\gins{P. Ginsparg, F. Quevedo, \np{385} (1992) 527.}
\lref\lykalw{S.P. DeAlwis, J. Lykken, \pl{269} (1991) 264.}
\lref\kolb{R. Holman, E.W. Kolb, S. Vadas, Y. Wang, \pl{250} (1990) 24.}
\lref\maeda{A.L. Berkin, K. Maeda, J. Yokoyama, \prl{65} (1990) 141.}
\lref\wise{L. Abbot, M. Wise, \np{244} (1984) 541.}
\lref\linde{B. Campbell, A. Linde, K.A. Olive, \np{355} (1991) 146.}
\lref\suss{A. Cooper, L. Susskind, L. Thorlacius, \np{363} (1991) 132.}
\lref\perr{V.A. Kostelecky, M.J. Perry, DAMTP-R92-40 (hepth 9302120), 12/92.}
\lref\das{S. Das, B. Sathiapalan, \prl{56} (1986) 2664.}
\lref\banks{T. Banks, Rutgers preprint RU-91-08.}
\lref\lastein{D. La, P.J. Steinhardt, \prl{62} (1989) 376.}
\lref\hol{R. Holman, E.W. Kolb, S.L. Vadas, Y. Wang, \pl{250} (1990) 24.}
\lref\giba{G. Gibbons, P. Townsend, \np{282} (1987) 71.}
\lref\gibb{G. Gibbons, K. Maeda, \np{298} (1988) 265.}
\lref\lid{A. Liddle, R. Moorhouse, A. Henriques, \np{311} (1988) 719.}
\lref\holb{R. Holman, E.W. Kolb, S. L. Vadas, Y. Wang, \prd{43} (1991) 995.}
\lref\freund{P.G.O. Freund, \np{209} (1982) 146.}
\lref\harvstr{J. Harvey, A. Strominger,  EFI-92-41.}
\lref\ant{I. Antoniadis, C. Bachas, J. Ellis, D.V. Nanopoulos, \pl{211} (1988)
393.}
\lref\barsa{I. Bars, D. Nemeschanski, \np{348} (1991) 89.}
\lref\barsb{I. Bars, K. Sfetsos, \pl{277} (1992) 269.}
\lref\hor{P. Horava, \pl{278} (1992) 101.}
\lref\witten{E. Witten, \prd{44} (1991) 314.}
\lref\gins{P. Ginsparg, F. Quevedo, \np{385} (1992) 527.}
\lref\gersh{D. Gershon, TAUP-1937-91.}
\lref\me{E. Raiten, FERMI-PUB-91-338-T.}
\lref\horne{J. Horne, G. Horowitz, \np{368} (1992) 444.}
\lref\bd{C. Brans, C.H. Dicke, \prd{124} (1961)  925.}
\lref\jord{P. Jordan, {\it Z. Phys.}{\bf 157} (1959) 112.}
\lref\gross{D. Gross, J.H. Sloan, \np{291} (1987) 41.}
\lref\kal{S. Kalara, K.A. Olive, \pl{218} (1989) 148.}
\lref\kalo{N. Kaloper, K.A. Olive, UMN-TH-1011/91.}
\lref\alb{A. Albrecht, P.J. Steinhardt, \prl{48} (1982) 1220.}
\lref\lindb{A.D. Linde, \pl{108} (1982) 389.}
\lref\la{D. La, P.J. Steinhardt, \prl{62} (1989) 376.}
\vfill
\eject
\newsec{Introduction}
For some time, it has been hoped that string theory would answer a number of
the central questions concerning low energy phenomena.  Initially, the
focus was the low energy gauge group, the number of generations, etc.
In part due to the increasing numbers of available ground states of string
theory (in conformal field theory, or in for example, the free fermion
approach), more recently attention has shifted to areas in which
gravitational dynamics plays a key role.  For example, string-theoretic and
string inspired models have played a significant role in the recent study of
information loss and Hawking radiation in two dimensions (see \harvstr\
for a recent review).

Another potential application of string theoretic methods lies in the physics
of the early universe.  It is by now well known that string theory contains
two principal differences from general relativity.  The first is the $\ap$
expansion of the equations of motion, which has Einstein's equations as the
leading term.  The second is the particular low energy spectrum of string
theory, which differs from more conventional models in the presense of the
dilaton, Kalb-Ramond field,
and tachyon (for bosonic strings).  As for the former,  the $\ap$
corrections should only be important for times of order $t_{PL}$ or less, and
furthermore, $R^2$ gravity can and has been studied without the input of
string theory.

As for the latter, the approach generally taken is to consider strings
propagating in a non-trivial background and demand conformal invariance.
This amounts to demanding vanishing beta functions, which in turn can be
interpreted as the field equations of an Einstein-like theory.  Generally,
these equations (and hence the corresponding field theory) are truncated at
leading order in $\ap$.  It is presumed that any solution of the field
equations corresponds to a well-defined ground state of string theory, i.e.,
a conformal field theory, but there is no proof of such a conjecture.
The alternative approach is to find an exact conformal field theory which
has the space-time interpretation of a cosmological model.  This approach
has yielded some success recently
through the use of non-compact Wess-Zumino-Witten models, \refs{\barsa ,
\barsb ,\hor ,\witten ,\gins ,\gersh ,\me ,\horne} , but
these solutions are generally not maximally symmetric and
often have singularities.  Furthermore a systematic
approach for finding cosmologically relevant solutions does not
seem to be available.

In the context of inflationary cosmology, for a number of reasons
the focus in the past has been on
the contribution of the dilaton \refs{\tsea ,\tsevaf ,\linde ,\suss ,
\kal ,\kalo}.
One reason is that the dilaton appears in
the action is a way similar to a Jordan-Brans-Dicke (JBD) field
\refs{\bd ,\jord}.
It should
be noted, however, as Campbell et al have discussed \linde ,  that the dilaton
cannot be identified with the standard Brans-Dicke field for any value of the
JBD parameter $\omega$.  It was also shown that including only the dilaton
and the metric did not yield an adequate inflationary model.
A second reason is the relation of the dilaton
to supersymmetry breaking.  Unfortunately, it is understood that the dilaton
potential is flat to all orders in string perturbation theory, and as usual,
the form of the nonperturbative potential is completely unknown, making it
very difficult to make definitive statements about dilaton induced inflation.

Considerably less attention has been paid to the contributions of the
tachyon.  In \suss , the tachyon was considered, but only in the context
of its propagation in the so-called linear dilaton background (see Section
2).
Another exception is its use in two dimensions \lykalw , where the
tachyon is actually massless.  However, as has been stressed recently by
Kostelecky and Perry \perr , the tachyon mass squared is negative in $d>2$
because one is expanding about a vanishing expectation value.  Using
the leading terms in the tachyon potential \das ,
they were able to show that the
tachyon in four dimensions has positive mass squared when expanded about the
minimum of its potential.

Kostelecky and Perry were primarily interested in the tachyon mass and the
nature of singularities in the metric under the freedom of making conformal
transformations using the dilaton.  Their analysis was done primarily in
the "sigma-model" frame (see Section 2), whereas most inflationary
models are analyzed in the Einstein frame.  By assuming a {\it constant}
tachyon background they were able to solve for the metric and dilaton,
assuming a $k=0$ Friedman-Robertson-Walker metric.

To discuss inflation, it is necessary to go beyond the assumption of a
constant tachyon, as it is by now well known that the dynamics of the
inflaton are crucial to any successful model.  That is what will be done
in this paper.  Of course, the equations of motion cannot then be solved
exactly, but that is not necessary.  Techniques now familiar in inflation
can be used to evaluate the essential quantities, such as the number of
e-foldings of the universe during inflation.  Various additional constraints
must also be satisfied in order to solve the smoothness, graceful exit,
and density perturbation problems.  Similar models have been investigated
in the Einstein frame in \maeda .

Section 2 will briefly review the background field equations and the
tachyon potential.  The constant tachyon solutions of \perr\ will also be
displayed.  In Section 3, the inflationary scenario will be discussed, and
in Section 4 the  various phenomenological constraints will be considered,
followed by some additional considerations in Section 5.
\vfill
\eject
\newsec{The $\beta$ function Equations and Tachyon Potential}

To leading order, the sub-Planckian energy physics of string theory can
be well described by propagation in the set of condensates of its massless
fields, namely the graviton
$g_{\mu \nu}$, antisymmetric tensor $B_{\mu \nu}$, dilaton $\p$, and tachyon
$T$.  For a world sheet $\Sigma$ with metric $\gamma_{ab}$
and local coordinates $\xi^a$, $a=1,2$, the location of the
string is described by spacetime coordinates $X^{\mu}(\xi^a)$, and
the action is
\eqn\back{I={{-1}\over {2\pi\ap}}\int d^2\xi\sqrt{\gamma}({1\over 2}\gamma^{ab}
\del_aX^{\mu}\del_bX^{\nu}g_{\mu \nu}
+{1\over 2}\epsilon^{ab}\del_aX^{\mu}\del_bX^{\nu}B_{\mu \nu}+
{1\over 2}\ap T -{1\over 4}\ap R^{(2)}\p ),}
where $R^{(2)}$ is the Ricci curvature of the world sheet.  It is this
curvature term which implies that the string coupling constant is
$\hat g=exp(\p /2)$.

As it stands, \back is not conformally invariant, as conformal invariance is
obviously broken by the tachyon and dilaton terms.  It can be restored
quantum mehcanically by requiring that
the beta functions for $g$, $B$, $T$ and
$\p$ vanish.  To one loop, these equations are
\eqn\grav{R_{\mu\nu}=\nabla_{\mu}\nabla_{\nu}\p+\nabla_{\mu}T\nabla_{\nu}T
 +{1\over 4}H_{\mu\nu\rho}H^{\mu\nu\rho}}
{\noindent for gravity}
\eqn\kr{\nabla_{\lambda}H^{\lambda\mu\nu}+(\nabla_{\lambda}\p )
H^{\lambda\mu\nu}==0}
{\noindent for the Kalb-Ramond field}
\eqn\tach{\nabla^2T+\nabla_{\mu}\p\nabla^{\mu}T=V'(T)}
{\noindent for the tachyon, and }
\eqn\dil{\hat c=R-(\del\p )^2-2\del^2\p-(\del T)^2-2V(T)-{1\over {12}}H^2}
{\noindent for the dilaton.
Here $H_{\mu\nu\rho}=3\del_{[\lambda}B_{\mu\nu ]}$ and $V'=
{{\del V}\over {\del T}}$, where $V$ is the tachyon potential which
shall be discussed shortly.  Also, $\hat c=2(d-26)/3\ap$, which
implicitly includes contributions from the conformal ghosts. }

These equations, in turn, can be derived from the following space-time
action
\eqn\stact{I_S={{-1}\over {2\kappa^2}}\int d^dx\sqrt{g}e^{\p}
(\hat c -R-(\del\p )^2+(\del T)^2+2V(T)+{1\over {12}}H^2),}
{\noindent where $2\kappa^2=8\pi G_N$ for $d=4$.  This is similar to
a JBD theory with $e^{\p}$ as the Brans-Dicke field.
This form of the action is
commonly known as the sigma-model frame because of its direct connection
to the sigma model action in \back .  By rescaling}
\eqn\conf{g_{\mu\nu}\rightarrow e^{-2\p/(d-2)}g_{\mu\nu}}
{\noindent the action can be put in the form}
\eqn\einstact{I_E={{-1}\over {2\kappa^2}}\int d^dx\sqrt{g}
((\hat c+2V(T))e^{-2\p/(d-2)}
 -R-{1\over {d-2}}(\del\p )^2+(\del T)^2
+{1\over {12}}e^{4\p/(d-2)}H^2).}

As stressed by Kostelecky and Perry, conformal invariance should guarantee
that one frame is as good as another, and the analysis below will generally
be conducted in the sigma-model frame.  In addition, there is of course
a Bianchi identity among the field equations.  One can therefore assume that
\dil\ is automatically satisfied if \grav -\tach\ are satisfied.

A simple solution to the field equation is the so-called linear dilaton
state, with the flat metric, vanishing Kalb-Ramond field and a dilaton
background of
\eqn\lindil{\p=\mu t,\ \mu^2={{2(d-26)}\over {3\ap}}.}
\noindent{Ignoring cubic and higher terms in the tachyon
potential (i.e., expanding about $T=0$), the tachyon must then satisfy}
\eqn\lindiltach{\nabla^2T+\mu\del_tT=-{4\over {\ap}}T}
\noindent{from which, after shifting T by a constant to eliminate the first
derivative term, one can read off the usual mass relation
$m^2={{2-d}\over {6\ap}}$.  The idea explored by Kostelecky and Perry \perr\ is
that the tachyon potential may have an alternative extremum
$T_0$ with positive $V''(T_0)$, thus leading to a positive mass and avoiding
the usual instability problems.  Since $V(T_0)$ is likely to be non-vanishing,
the resulting solutions are likely to behave as de Sitter or anti-de Sitter
space, so they assumed a k=0 Friedman-Robertson-Walker (FRW) metric}
\eqn\frw{ds^2=-dt^2+a(t)^2(dx_1^2+\cdots dx_{d-1}^2).}
{\noindent Flat spatial sections are favored, since the false vacuum $T=0$
can then evolve into the true vacuum without any topology change.  With this
assumption, the field equations can be written}
\eqn\grava{(d-1){{\ddot a}\over a}+\ddot\p+\dot T^2=0}
\eqn\gravb{a\ddot a+(d-2)\dot a^2+a\dot a\dot\p=0}
\eqn\tachfrw{\ddot T+(d-1){{\dot a}\over a}\dot T+\dot \p\dot T+V'(T)=0}
\eqn\dilfrw{\hat c=\dot \p^2+\ddot\p+(d-1){{\dot a}\over a}\dot\p-2V(T)}
{\noindent where \dilfrw\ has already been used to simplify \grava .}

The solutions of Kostelecky and Perry are found by letting $T=T_0$ where
$V'(T_0)=0$, i.e., the tachyon sits at its minimum.  Letting
$h={\rm ln}(\sqrt{\beta}\dot a/a)$, $\p$ can be eliminated from the
remaining equations,
which then reduce to
\eqn\kpeqn{\beta \ddot h=(d-1)e^{2h}.}

{\noindent  Note that the Hubble paramter $H=\dot a/a=e^h/\sqrt{\beta}$.
Kostelecky and Perry chose $\beta =\ap$, which is the natural
unit of time, but that is obviously not required, since $\ap$ does not appear
in \grava -\dilfrw .  In fact, it will be essential in what follows
that $\beta$ be different from $\ap$ (see the next section).
\kpeqn\ can be integrated once to yield}

\eqn\kpeqnb{\dot h^2={{d-1}\over {\beta}} e^{2h}+k}

{\noindent where $k$ is an integration constant.  For $k=0$ the solution is}
\eqn\kpsola{a=a_0t^{1/\lambda},\ \p=\p_0 +(1-\lambda ){\rm ln}(t/\sqrt{\beta})}
{\noindent where $\lambda =\pm (d-1)^{1/2}$
(only the positive square root will be needed here).
For $k>0$ the solution is}

\eqn\kpsolb{a(t)=a_0({\rm tanh}(\lambda t\sqrt{k}/2))^{1/\lambda}}
\eqn\kpsolc{\p(t)=\p_0+(1+\lambda) {\rm ln (cosh}(\l\sqrt{k} t/2))
+(1-\l){\rm ln (sinh}(\l \sqrt{k} t/2)).}
{\noindent Note that the two solutions are equivalent in the $t\rightarrow 0$
limit.  For $k<0$, solutions similar to \kpsolb\ exist, with tanh replaced by
tan, etc, but there, the solution diverges in a finite amount of time, which
is assumed to be unphysical. }

Thus far, no assumptions concerning the form of the tachyon potential
have been made.  This has been evaluated in a number of references, see
for example \suss\ and \das , in string field theory.  These calculations
are perturbative in the tachyon, but sum all orders of the loop expansion.
The result is
\eqn\tachpota{\hat V(\hat T)=-{2\over {\ap}}\hat T^2+
{g\over {3!}}\hat T^3+\cdots .}
{\noindent where $g$ is the tree-level three-tachyon coupling defined at
zero momentum.  Now,  in \tachpota , $\hat T$ has mass dimensions $12$ and
the tachyon potential has mass dimension $26$, as appropriate for the string
field theory calculation, which is conducted in the critical dimension
$d=26$.  Kostelecky and Perry chose to work with a dimensionless tachyon and
a tachyon potential of mass dimension $2$ by rescaling}
\eqn\rescala{\tilde T=\kappa g\ap\hat T,\ V(\tilde T)=\kappa^2g^2\ap^2
\hat V(\hat T)}
\eqn\tachpotb{\tilde V(\tilde T)=-{2\over {\ap}}\tilde T^2+
{1\over {3!\kappa \ap}}\tilde T^3+\cdots ,}
{\noindent where $\kappa$ is a
dimensionless constant.  In the cosmological context, it
seems more natural to choose a dimension one tachyon and dimension four
potential, so instead the rescaling}
\eqn\rescalb{T=\kappa g \ap^{1/2}\hat T,\ V(T)=\kappa^2g^2\ap \hat V(\hat T)}
\eqn\tachpotc{V(T)=-{2\over {\ap}}T^2+{1\over {3!\kappa\sqrt{\ap}}}T^3}
{\noindent will be used here.\footnote{$^\dagger$}{Note:  This choice of
rescaling induces various factors of $8\pi G_N$ which will be suppressed
below.}}

The freedom to choose $\kappa$ reflects the fact that only the $T^2$ term
in the
potential is universal.  The original calculations
\refs{\suss ,\das} indicate that
$\kappa =1/4$, but the result is dependent on the regularization proceedure.
It is presumed that varying $\kappa$ would generally induce higher derivative
terms \banks .  Since such terms are generally unimportant during
inflationary epochs, it seems plausible to leave $\kappa$ as a free
parameter for now.

\newsec{The Inflationary Scenario}

It is clear that the solution in equation \kpsola\ is unsatisfactory for
inflation, as it is well known \wise\ that power law inflation $a\sim t^p$
requires that the power $p$ be of order $10$.  Instead, one should proceed
as in more standard inflationary models \refs{\la ,\alb ,\lindb}.
There, one assumes that the
inflaton, here the tachyon, undergoes a \lq\lq slow roll'' towards its true
minimum.  It is during this slow roll that the universe inflates.

First, it is useful to eliminate $\p$ from \grava\ and \tachfrw\ by using
\gravb\ .  This yields
\eqn\reducegr{\dot T^2=\ddot h -(d-1) e^{2h}/\beta}
\eqn\reducetach{\ddot T -\dot h\dot T+V'(T)=0.}
{\noindent The physical picture is to suppose that for small $t$, the
tachyon is near the false vacuum of zero, and that the metric and dilaton
have the form as in say \kpsola\ .  For small $t$, $\dot h<<0$, so that
from \reducetach\ it is clear that the tachyon is overdamped and deviations
from $T=0$ strongly suppressed.  As time goes on $|\dot h|$ decreases, and
presumably a quantum fluctuation starts the tachyon rolling towards its
true minimum.}

The standard proceedure is to then assume that both
the $\ddot T$ term in
\reducetach\ and the $\dot T^2$ term in \reducegr\ are small.  However, one
only ignores $\dot T^2$ relative to $V(T)$ during the slow roll, but
in contrast to the usual case of minimally coupled scalars, $V(T)$
does not appear in \reducegr\ , so this term must be kept (otherwise,
\reducegr\ is the same as \kpeqn\ and its solutions known!).  The question
as to whether $\ddot T$ is small will be considered shortly, but to the
extent that it is small, $T$ can be approximated as a linear function of $t$,
i.e.,
\eqn\tlin{T(t)=C t+T_0,}
{\noindent while the tachyon equation is}
\eqn\tachapprox{\dot h\dot T=V'(T),}
{\noindent where $C$ and $T_0$ are constants which will be estimated below.
Since then $\dot T=C$, \reducegr\ can still be integrated once, yielding}
\eqn\heqn{\dot h^2-(d-1)e^{2h}/\beta-2C^2 h=k_2}
{\noindent where $k_2$ is again an integration constant.  Equation \heqn\
does not appear to be solveable in closed form.  As a somewhat crude
approximation, let us assume that the exponential in \heqn\ can be
approximated by the first two terms in its power series.  Then \heqn\ is
approximately}

\eqn\happrox{\dot h^2=C_1+C_2 h}
\eqn\conectwo{C_1=(d-1)/\beta +k_2,\ C_2=2(d-1)/\beta +2C^2.}
{\noindent This can now be integrated, and the solution is}
\eqn\hsol{h(t)=h_0\pm t\sqrt{C_2h_0+C_1}+C_2t^2/4,}
{\noindent where $t=0$ has been taken as the start of the inflationary
epoch and $h_0$ is the initial condition on $h$.  It is clear from
\tachapprox\ , since $\dot T>0$, $V'(T)<0$, that $\dot h$ must be less than
zero.  This necessitates taking the minus sign in \hsol , and more
importantly implies that the inflationary period (or rather, the period
during which the above approximations are valid) must end when $\dot h\sim 0$.
This occurs when }
\eqn\endinf{t=t_0=2\sqrt{h_0+C_1/C_2}=2\dot h_0/C_2.}

At this point, given $h(t)$, the number of e-foldings through which the
universe passes during the inflationary epoch can be evaluated.  This is
given by
\eqn\efolds{N=\int_0^{t_0}H dt={1\over {\sqrt{\beta}}}\int_0^{t_0}e^h}
{\noindent which, after some simple manipulations, can be expressed as}
\eqn\nefolds{N=4 {{e^{-C_1/C_2}}\over {\sqrt{\beta C_2}}}
\int_0^{\sqrt{h_0+C_1/C_2}}e^{x^2}dx.}
{\noindent In general, $\beta C_2$ is of order one, and some numerical
evaluation then implies that for $C_1\sim C_2$, that a value of $h_0$
of about $5$ is sufficient for the 50-60 e-folds which are necessary to
solve the flatness and smoothness problems.  Of course, $h_0=5$ means that
the approximation made in \happrox\ is not well justified.  But for
$C_1\sim C_2$, it can be demonstrated numerically that the associated errors
are not too large, and the qualitative picture is unchanged.  In any case,
it is clear that the exponential factor tends to make $h$ larger
(for fixed $t$), which in
turn make $e^h$ much larger which therefore leads to an even
larger expansion of the universe (even thought the duration of the
expansion is slightly smaller).}

When $\dot h>0$, the $\ddot T$ term cannot be neglected, and the tachyon
accelerates
towards its true minimum.  It will then oscillate about this minimum, thereby
reheating the universe.  To discuss this epoch, consider again equation
\tachfrw\ in $d=4$.  Then $V'(T)$ can be set to zero, since the tachyon is
near its minimum.  Therefore, in this region, \tachfrw\ has the standard
form of a scalar wave equation in an expanding universe with decay constant
equal to $\dot \p$.  There is one difference, however.  From the action
$I_S$ in equation \stact\ , one easily determines that the energy momentum
tensor for the tachyon has an additional factor of $e^{\p}$, i.e., its
energy density and pressure are given by
\eqn\rhoT{\rho_T=e^{\p}(\dot T^2/2+V(T))}
\eqn\pT{p_T=e^{\p}(\dot T^2/2-V(T))}
{\noindent After multiplying \tachfrw\ by $e^{\p}\dot T$, it can be rewritten
as}
\eqn\tachdec{\dot\rho_T+3H\rho_T=-\dot\p\rho_T,}
{\noindent where, as usual, the virial theorem for the harmonic oscillator
has been used, and the potential term in \rhoT\ ignored (since it is roughly
constant during the oscillations about the minimum).  Then \tachdec\ can
be integrated to yield}
\eqn\tachrhodec{\rho_T=M^4(a/a_0)^{-3}e^{\p_0-\p},}
{\noindent  where $M$ is a constant of integration.  For example, if
it is assumed that the solution is roughly similar to
the constant $T$ solution \kpsolb\ , where $\p_0-\p$ varies
as $-\lambda\sqrt{k}t$ for large t, then the tachyon energy density decays
in the usual fashion.  In particular, for sufficiently large $k$, the
decay and therefore the reheating of the universe is quite rapid, and the
reheat temperature on general grounds is
basically equal to $M$, i.e., the vacuum energy is converted directly into
radiation.  More generally, during the oscillation epoch, the maximum
temperature is
rougly given by \kt\ }
\eqn\reheatmax{T_{RH}\sim T_{max}\sim M^{1/2}(\lambda \sqrt{k} m_{PL})^{1/4}.}
{\noindent The temperature then decreases during
the oscillation phase, and the temperature
at the beginning of the radiation-dominated epoch is roughly \kt\ }
\eqn\reheattemp{T_{RH}\sim (m_{PL}\lambda \sqrt{k})^{1/2}.}

Before considering the remaining inflationary constraints, it is appropriate
at this point to determine the constant $C$ in \tlin .  Consider the limit
of small $t$
(as measured from the start of the inflationary epoch),
for which \tlin\ should be a good approximation.  It will also be assumed
that in this limit $T$ itself is small.  Substituting \tlin\ and \hsol\
into \tachapprox , ignoring quadratic and higher order terms in $V'(T)$,
and assuming that $C_1\sim C_2$, one finds
\eqn\cmatch{C^2=-{{d-1}\over {\beta}}+ \sqrt{({{d-1}\over {\beta}})^2+
{{T_0^2}\over {2\ap^2}}}.}
{\noindent where $h_0\sim 5$ has been assumed.
The physical interpretation of \cmatch\ is relatively clear;
smaller values of $T_0$ correspond to the tachyon starting very near its
(false) vacuum, where the potential is quite flat so that the starting
velocity is correspondingly small.}

\newsec{Inflationary Constraints}

A number of issues remain to be answered before the scenario described above
can be considered self-consistent.  First, there were a number of
approximations which must be justified for some non-vanishing range of
parameters.  The most important of these was the slow-roll approximation.
The general condition which should satisfied for the slow-roll to be a
good approximation is \kt
\eqn\slowroll{|V''(T)| < 9H^2.}
{\noindent Assuming that before inflation the metric is behaving as in
\kpsola , then for small $T$, \slowroll\ gives a condition on the
starting time of the inflationary epoch, namely}
\eqn\slowrcond{t < \sqrt{\ap}/2.}
{\noindent Now, since it was shown above that $h_0>5$ is necessary for
sufficient inflationary expansion, the parameter $\beta$ can be determined.
This implies}
\eqn\betadet{\beta /\ap > 10^4.}

Another possible
phenomenological constraint is the graceful exit problem.  Simply
put, the graceful exit problem is the apparant conflict between adequate
inflation, which requires a small bubble nucleation rate, and percolation,
which requires a fast bubble nucleation rate so that the phase transition
goes to completion.  This has been considered in \hol\ for the general
case of extended inflation \lastein\ , and applied to string theoretic
models in \linde .  The results of \hol\ imply that the nucleation rate
$\bar\lambda$ varies as
\eqn\nucl{\bar\lambda = \hat A {\rm exp}(2\p-B_0 e^{\p})}
{\noindent where $\hat A$ and $B_0$ are
positive constants which are computed
in \hol\ but which do not depend on the inflationary dynamics.  It normally
suffices to show that the nucleation rate is a non-decreasing as a function
of time.  From \gravb\ it is trivially shown that}
\eqn\dilhrel{\dot\p =-\dot h-(d-1)e^h/\sqrt{\beta} ,}
{\noindent which can be integrated using \efolds\ to determine the change
in the dilaton $\delta\p =\p (t_0)-\p_0$ during inflation}
\eqn\dilinf{\delta\p =h_0-h(t_0)-N (d-1)\sim -200,}
{\noindent for $N\sim 60$.  Therefore, the dominant factor in \nucl , namely
the second term inside the exponential, leads to a non-decreasing nucleation
rate and the phase transition should proceed to completion.}

The last phenomenological condition that will be considered here is that
of density perturbations.  Of course, these cannot be too large in view
of the homogeneity of the microwave background, nor too small in view of
the recent COBE results.  The now standard parameter which one must
evaluate is $(\delta\rho /\rho )_H$ where
$\rho$ is the density in a particular wavenumber, and where $H$
in the subscript refers
to the time when that perturbation reenters the horizon \kt .  Actually,
the standard calculation is to compute a gauge invariant quantity which
equals
$\delta\rho /(\rho +p)$ at horizon crossing, and which equals
$(\delta\rho /\rho )_H$ up to a factor of order unity.  This can then
be computed when the perturbation leaves the horizon during inflation.
One then takes $\delta\rho =\delta T e^{\p}{{\del V(T)}\over {\del T}}$ and
$\delta T=H/(2\pi)$.  The latter statement is calculated for a minimally
coupled scalar.  If the dilaton were constant during inflation, we could
roughly speaking compensate for the tachyon-dilaton coupling by including
an additional factor of $e^{\p /2}$.  One then must use the
equation of motion for the tachyon during inflation (it is at this point
where the unusual dilaton couplings come into play), i.e., equation
\tachapprox .  One then has
\eqn\deltarho{\delta\rho /\rho \sim {{e^{+\p /2}\dot H}\over {\dot T}}.}
{\noindent For perturbations which leave the horizon near the end of
inflation (i.e., small scale perturbations), the right hand side of
\deltarho\ is small because $\dot H=H\dot h\sim 0$.  However, for scales
which leave the horizon in the beginning of inflation, the Hubble parameter
is large (since $h_0\sim 5$), and the result in \slowrcond\ implies that
$\dot h$ is large as well.  This
can be used to put a constraint on the initial value of
the dilaton $\p_0$.  Alternatively, this potential problem can be removed
if one assumes that $\sqrt{\ap}\gg t_{PL}$, but that is somewhat unnatural
in the context of string theory.}

Similar models have been discussed by Berkin et al \maeda\ in the Einstein
frame.  In those models, because of the power law behavior of the scale
factor, the constraint due to density perturbations are found to be somewhat
weaker than in conventional inflation models.  This suggests that a more
complete treatment of the tachyon perturbations should not cause a serious
problem.

\newsec{Discussion}

While a complete examination of the parameter space of tachyon-induced
inflation has not been attempted here, it has been shown that
cosmologically viable models do exist.  In fact, it is gratifying that
no dimensionless parameter of order $10^{-15}$ arose in the construction
or in the constraints, in sharp contrast to most conventional
inflationary models (though several paramters of order $100$ have arisen).

One difference between the model presented here and most inflationary
models is the use here of the \lq\lq string'' frame, often referred to as
the Jordan frame in cosmology.   When the dilaton is constant, the change
between frames is quite trivial (merely corresponding to a linear
rescaling of $t$),
but when the dilaton is dynamical the
correspondence between frames is often rather obscure.  This is primarily
due to the fact that the rescaling \conf\ also requires a redefinition
of \lq\lq time'' to maintain the Friedman-Robertson-Walker form of the
metric.  This has recently been emphasized by Tseytlin \tsea , who showed
that the scale factor $a_S$ and $a_E$ in the two frames are related by
\eqn\scalfr{{\rm ln}(a_E) ={\rm ln}(a_S)
-{2\over {d-2}}\p .}
{\noindent For example, it is possible for the metric to be static in
one frame and linearly rising in the other.
In \perr\ it is argued
that neither frame should be considered as more fundamental
than the other.  In fact, they make an even stronger suggestion, namely
that singularities in the metric should be considered essentially benign
as long as a frame can be chosen in which the singularity is not present
(in addition to the usual way of avoiding singularities by diffeomorphisms
when the singularity is merely a coordinate singularity).  This criteria
then determines the preferred choice of frame.
While
singularities are not the issue here, the same principle should apply,
i.e., the frame can be chosen so as to fit the requirements of inflation.
One should also consider the fact that phyiscal scales, such as
$\Lambda_{QCD}$, vary in time depending on the chosen frame, and the expansion
of the universe should in principle be measure against those scales, and
not just the string scale \linde .}

There are a number of possible extensions of the present work.  For example,
the anti-symmetric tensor has thus far been ignored.  This has been
considered in previous work which did not include the tachyon, see for
example \refs{\tsea ,\giba ,\gibb ,\lid ,\ant ,
\freund ,\holb} .  The solution in \ant\ can in
fact be related to a Wess-Zumino-Witten model conformal field theory, while
other solutions have arisen in the context of Kaluza-Klein theories.  In
\tsea\ it is claimed that the presence of the Kalb-Ramond field does not
greatly change the behavior of the system.  As mentioned in \perr ,  it
would be quite interesting to consider solutions to the background field
equations for the heterotic string, since that is a much more viable
candidate for a grand unified model, though of course that would involve
additional fields which would complicate the analysis.  The effective
action up to quartic order, has been computed for the heterotic string in
\gross .

A final important question concerns the necessity of adding higher order
terms in the $\ap$ expansion.  Indeed, if
$\sqrt{\ap}\sim t_{PL}$, then equation \slowrcond\ would suggest the necessity
of adding
such terms.  Such higher order terms would change the initial conditions
for inflation, in particular the relationship between $h_0$ and $\dot h_0$
which were used implicitly in the last section.  But the dynamics of
the inflationary epoch itself, when such terms should be small, should
remain unchanged.

\vfill
\eject
\listrefs
\bye